\documentclass[superscriptaddress, twocolumn, amsmath,amssymb, aps, prl]{revtex4-1}

\usepackage{graphicx}
\usepackage{dcolumn}
\usepackage{bm}
\usepackage{hyperref}
\usepackage{xcolor}
\usepackage{fancyhdr}

\begin{document}

\newcommand{\RHheaderline}{\textsf{P3H-23-101, TTK-23-34, MPP-2023-286, IFJPAN-IV-2023-6}}
\fancypagestyle{firstpage}
{
  \renewcommand{\headrulewidth}{0pt}
  \fancyhead[R]{\RHheaderline}
}

\title{Top-Bottom Interference Contribution to Fully-Inclusive Higgs Production}

\author{Micha\l{ }Czakon}%
 \affiliation{Institute for Theoretical Particle Physics and Cosmology, RWTH Aachen University, 52056 Aachen, Germany.}
\author{Felix Eschment}%
 \affiliation{Institute for Theoretical Particle Physics and Cosmology, RWTH Aachen University, 52056 Aachen, Germany.}%
\author{Marco Niggetiedt}
 \affiliation{Max-Planck-Institut f\"ur Physik,\\ Boltzmannstra{\ss}e 8, 85748 Garching, Germany.}%
\author{Rene Poncelet}
 \affiliation{The Henryk Niewodnicza\'{n}ski Institute of Nuclear Physics,\\ ul.\ Radzikowskiego 152, 31-342 Krakow, Poland}
\author{Tom Schellenberger}%
 \affiliation{Institute for Theoretical Particle Physics and Cosmology, RWTH Aachen University, 52056 Aachen, Germany.}%

\date{\today}

\begin{abstract}
We evaluate the top-bottom interference contribution to the fully-inclusive Higgs production cross section at next-to-next-to-leading order in QCD. Although bottom-quark-mass effects are power-suppressed, the accuracy of state-of-the-art theory predictions makes an exact determination of this effect indispensable. The total effect of the interference at 13~TeV is $-1.99(1)^{+0.30}_{-0.15}$~pb, while the pure $\mathcal{O}(\alpha_s^4)$ correction is 0.43~pb. With this result, we address one of the leading theory uncertainties of the cross section.
\end{abstract}

\maketitle
\thispagestyle{firstpage}


\section{Introduction}
The discovery of the Higgs boson at the Large Hadron Collider (LHC) by the ATLAS \cite{ATLAS:2012yve} and CMS \cite{CMS:2012qbp} collaborations was a major breakthrough for the field of particle physics. A comprehensive understanding of the newly found boson and all its properties is pivotal to investigating possible new physics scenarios. Among the many aspects of Higgs physics phenomenology, the production cross section is central. This quantity is dominated ($\sim\! 90\%$ at 13~TeV center-of-mass energy) by the gluon-fusion channel $gg\to H$, where two gluons couple to the Higgs boson via a quark triangle loop. Thus, extensive efforts have been devoted to determining the cross section of this channel precisely. One direction towards better accuracy is to include higher orders in the strong coupling constant $\alpha_s$ in the calculation. Since this comes with the difficulty of calculating amplitudes with more and more loops, a commonly used approximation is the heavy-top limit (HTL) given by $m_t\to \infty$. This introduces an effective coupling between gluons and the Higgs, transforming the loop-induced $gg\to H$ process into a tree-level process at leading order (LO) in the effective theory. In the latter, calculations are significantly simplified by reducing the loop order by one and eliminating one of the scales. This has enabled calculations of the Higgs production cross section in gluon-gluon fusion at next-to-leading~\cite{Dawson:1990zj,Djouadi:1991tka}, next-to-next-to-leading~\cite{Ravindran:2002dc, Harlander:2002wh, Anastasiou:2002yz}, and next-to-next-to-next-to-leading~\cite{Anastasiou:2015vya,Mistlberger:2018etf} order (NLO, NNLO, $\text{N}^3$LO). These works have shown that Higgs production receives large corrections from higher orders. Notably, the $\mathcal{O}(\alpha_s^3)$ cross-section contribution surpasses the LO prediction, while NNLO corrections remain substantial, contributing roughly $20\%$ to the total rate. $\text{N}^3$LO corrections ($\sim\! 3\%$) reduce the scale uncertainties to the percent level. Due to this remarkable accuracy, it becomes important to examine potential other sources of uncertainty carefully. A comprehensive analysis of the status of the cross-section calculation has been performed in Ref.~\cite{LHCHiggsCrossSectionWorkingGroup:2016ypw}. While uncertainties stemming from parton distribution functions (PDFs) and their unavailability at $\text{N}^3$LO remain until today, others have been reduced by calculating the $\text{N}^3$LO contribution without a truncated threshold expansion~\cite{Mistlberger:2018etf} and by including the mixed QCD-electroweak corrections~\cite{Bonetti:2018ukf,Bonetti:2020hqh, Becchetti:2020wof,Bonetti:2022lrk}. All remaining sources of theory uncertainty are related to effects from finite quark masses. The elimination of these uncertainties requires calculations in full QCD beyond the HTL. At LO~\cite{Georgi:1977gs} and NLO~\cite{Graudenz:1992pv}, this has been achieved by explicit calculation a long time ago. For higher orders (HO), it turns out that it is a good approximation to multiply the HTL result with a rescaling factor containing the LO impact of a heavy top quark---this procedure is also known as Higgs Effective Field Theory (HEFT):
\begin{equation}
    \label{eq:HEFT}
    \sigma^\text{HO}_\text{HEFT}=\frac{\sigma^\text{LO}_t}{\sigma^\text{LO}_\text{HTL}}\,\sigma^\text{HO}_\text{HTL}\approx 1.065\times \sigma^\text{HO}_\text{HTL},
\end{equation}
where the numerical value was calculated using $m_H=125$~GeV and $m_t=173.055$~GeV. Still, missing finite-quark-mass effects at NNLO remain a source of uncertainty. In particular, two contributions were each estimated to amount to $\sim\! 1\%$ in Ref.~\cite{LHCHiggsCrossSectionWorkingGroup:2016ypw}: the exact dependence on the top mass in all diagrams where the top quark remains the only quark which couples to the Higgs, and interference effects between such top-quark diagrams and diagrams containing a bottom or charm loop. These finite-quark-mass effects are generally power-suppressed: In the case of finite top-mass effects, the cross section contributions are suppressed by powers of $m_H^2/4m_t^2$ close to the production threshold. For bottom quarks, on the other hand, they are suppressed by $m_b^2/m_H^2$ independently of partonic center-of-mass energy. The latter suppression originates partly from the Yukawa coupling constant and partly from the scalar nature of the interaction vertex, which requires a helicity flip of the quark. Helicity conservation then demands that the cross section vanishes as $m_b \to 0$, resulting in an overall suppression of $m_b^2$. However, the cross section also receives strong logarithmic enhancements of the form $\log^2 \! \left(m_b^2/m_H^2 \right)$, rendering the total interference quite sizeable. Charm-quark contributions receive an analogous suppression, which is however even stronger due to the smaller mass value. Top- and bottom-quark-mass effects are hence viewed to be among the leading theory uncertainties. Recently, the production cross section was computed with full top-mass dependence at NNLO~\cite{Czakon:2021yub}, confirming the expected size of the top-mass effects. So far, predictions of the fully inclusive top-bottom interference cross section are still lacking. A threshold expansion around $\hat{s} = m_H$ at NLO with next-to-next-to-leading logarithmic resummation in the HTL~\cite{Anastasiou:2020vkr} has been used to estimate the interference to contribute -2.18~pb at NNLO. On top of that, differential cross-section analyses for Higgs+jet production at NLO \cite{Grazzini:2013mca, Melnikov:2016emg, Lindert:2017pky,Caola:2018zye, Bonciani:2022jmb} have become available and confirmed the significance of bottom-quark effects, especially in the low-$p_T$ region. In these studies, the transverse momentum of the Higgs serves as an infrared regulator. In the case of the fully inclusive cross section, the possibility of a vanishing Higgs transverse momentum has to be taken into account. This necessitates the inclusion of double-virtual amplitudes and also a numerically stable implementation of real-radiation amplitudes in infrared-sensitive regions of the phase space.

In this letter, we address these points, which enables us to compute the interference effect for the fully-inclusive Higgs production cross section at NNLO.

\section{Calculation}
In order to compare the size of the top-bottom interference effects with the known contributions to the total rate, we adopt the 5-flavor scheme, wherein the bottom quark is included in the proton PDFs. In this scheme, the bottom quark is treated as massless, in contrast to the 4-flavor scheme, where the bottom quark is absent from the initial state and treated as a massive particle. The non-vanishing mass of the bottom quark cannot be completely disregarded, since, as we already pointed out, the interference effects are power suppressed by a factor proportional to $m_b^2$. Hence, we treat the bottom quark as a massive particle, but only inside closed fermion loops that couple to the Higgs. This approach assures that the resulting amplitudes are gauge invariant. The reason can be easily understood, as the calculation is formally equivalent to introducing a replica bottom quark which is however absent in the PDFs and decoupled from the running of $\alpha_s$. The Yukawa coupling to the Higgs is an arbitrary parameter from the perspective of QCD, therefore, the set of diagrams in which the replica bottom quark couples to the Higgs must be independently gauge invariant.

To compute the cross section, we perform the phase-space integration of the relevant amplitudes using Monte Carlo techniques. For our purposes, we need amplitudes with both real and virtual corrections to Higgs production in gluon fusion. Namely, we require one-loop double-real, two-loop real-virtual and three-loop double-virtual amplitudes. The double-real amplitudes are the same as in the case of a single heavy quark, except that the involved integrals are now also evaluated above the quark production threshold $m_H^2 > 4 m_q^2$. The amplitudes have been calculated in Ref.~\cite{Budge:2020oyl} and we use their implementation inside \texttt{MCFM} \cite{Campbell:2019dru, Campbell:1999ah, Campbell:2011bn}, which itself uses \texttt{QCDLoop} \cite{Ellis:2007qk, Carrazza:2016gav} to calculate the scalar integrals. Starting with the two-loop real-virtual corrections, the amplitude receives corrections from mixed quark loops, i.e.~diagrams with both a top and a bottom quark loop. Fortunately, the resulting integrals can always be factorized into separate standard one-loop integrals. The main challenge arises from two-loop amplitudes with only a single fermion loop. Here we follow the strategy outlined in Ref.~\cite{Czakon:2021yub}, which itself is based on Refs.~\cite{Boughezal:2007ny, Czakon:2008zk}. The general idea is to reduce the amplitude to a set of master integrals, which is done with the help of the public software \texttt{Kira}$\oplus$\texttt{FireFly} \cite{Maierhofer:2017gsa, Maierhofer:2018gpa, Klappert:2019emp, Klappert:2020aqs, Klappert:2020nbg}, and then derive a set of coupled differential equations. The master integrals are functions of three variables which are chosen to be 
\begin{equation}
\xi = 1 - \frac{m_H^2}{\hat{s}}, \quad \lambda = \frac{\hat{t}}{\hat{t} + \hat{u}}, \quad \text{and} \quad x = \frac{m_q^2}{\hat{s}},
\end{equation}
where $\hat{s}, \hat{t}$ and $\hat{u}$ are the usual partonic Mandelstam variables of the process $g g \to g H $ or $q \bar{q} \to g H$, and $m_q$ and $m_H$ refer to the masses of the heavy quark and the Higgs boson. The amplitudes for these processes are symmetric with respect to $\lambda \to 1 - \lambda$, and all other relevant amplitudes are related by crossing. The boundary conditions are determined in the limit of large $x$ by a large-mass expansion as described in Ref.~\cite{Czakon:2020vql}. Subsequently, the solution on the boundary is transported to physical values of $x$ using the differential equation for a fixed value of $\lambda$ close to the symmetry axis $1/2$ and multiple values of $\xi$. In the next step, the differential equation is solved in $\lambda$ to map out the entire physical parameter space, such that we obtain a dense grid for the amplitudes, which can then be used for the Monte Carlo integration. 

When performing this integration, it is necessary to sample phase-space points which differ from the grid points. The required interpolation is difficult at the boundaries if infrared divergences are present. Hence, we remove them by subtracting the rescaled amplitudes in the HTL from the full result. The rescaling parameter $r$ is chosen specifically so that the LO cross sections in the effective and the full theory match exactly:
\begin{equation}
    \label{eq:rescaling-tb}
    r=\frac{\sigma_{t\times b}^\text{LO}}{\sigma^\text{LO}_\text{HTL}}\approx -0.129,
\end{equation}
with $m_b=4.779$~GeV additionally to the values used for Eq.~\eqref{eq:HEFT}. This procedure assures that infrared divergences arise solely from tree-level soft or collinear radiation, which we subtract with the help of the corresponding splitting function as well. Afterwards, the regulated grid is interpolated using cubic splines, whereas the subtracted terms can be added back at arbitrary phase-space points since they are known as compact analytic expressions. Notice the resemblance of the rescaling in Eq.~\eqref{eq:rescaling-tb} to that in Eq.~\eqref{eq:HEFT}. While the latter employs a rescaling parameter close to one and is commonly used to correct for finite top-quark mass effects, the rescaling performed here should only be understood as a computational trick, which is underlined by the negative value of~$r$.

The three-loop virtual corrections are exclusively gluon-initiated since the Higgs-quark form factor is zero to all orders of perturbation theory for massless quarks.
The Higgs-gluon form factor, on the other hand, receives genuine corrections. Contributions containing only a single type of heavy quarks, i.e.~either a top or a bottom quark, have been evaluated both above and below the quark production threshold $2 m_q$ in Ref.~\cite{Czakon:2020vql}. Results for mixed heavy-quark contributions have been published recently in Ref.~\cite{Niggetiedt:2023uyk}. In both cases, the authors first employed a high-energy and/or large-mass expansion of the integrals. Afterwards, they used differential equations to extend the expansion to very high orders such that the amplitudes can be calculated precisely.

All contributions have been included in \texttt{Stripper}, the \texttt{C++} implementation of the sector-improved residue subtraction scheme \cite{Czakon:2010td,Czakon:2014oma,Czakon:2019tmo}, in order to correctly deal with all their infrared divergences. The cancellation of infrared divergences is checked by verifying that all poles in the dimensional regulator $\epsilon$ are compatible with zero within their Monte Carlo uncertainty.

The masses of the heavy quarks are renormalized in the on-shell scheme, while the strong coupling constant is renormalized in the $\overline{\text{MS}}$ scheme with five light quark flavors. The Lehmann-Symanzik-Zimmermann (LSZ) constant for the gluon field is non-trivial already at the one-loop order \cite{Czakon:2007wk}. The LSZ constants for the massless quark fields are non-trivial starting at two-loops, which makes them irrelevant for the present calculation. Since the bottom quark is assumed to be massless whenever the fermion loop is not coupled to the Higgs boson, it does not contribute to the gluon LSZ constant, and no decoupling constant needs to be introduced.

\section{Results}
\begin{table*}
\begin{ruledtabular}
\caption{Effects of interference of bottom- and top-quark amplitudes on Higgs production in the gluon-fusion channel at LHC @ 7, 8, 13, 13.6, and 14~TeV. The results are obtained with the \texttt{NNPDF31\_nnlo\_as\_0118} \cite{NNPDF:2017mvq} PDF set with a Higgs mass of $m_H = 125$~GeV and quark masses of $m_t = \sqrt{23/12}\, m_H \approx 173.055$~GeV and $m_b = \sqrt{1/684}\, m_H \approx 4.779$~GeV. In the second-to-last column, the bottom Yukawa coupling $Y_b=\sqrt{2}\, m_b/v$ is renormalized in the $\overline{\text{MS}}$ scheme, where $m_{b,\overline{\text{MS}}}(m_{b,\overline{\text{MS}}}) = 4.18$~GeV. The calculation was performed at a central scale of $\mu_F = \mu_R = m_H/2$. The scale uncertainties were determined with seven-point variation. The same scale setting was used in the numerator and denominator in the ratio presented in the last column. The HEFT values have been obtained with \texttt{SusHi}~\cite{Harlander:2012pb,Harlander:2016hcx}.}
\label{tab:results}
\centering
\begin{tabular}{cccccc}
Order  &  $\sigma_\text{HEFT}$ [pb] & $(\sigma_t - \sigma_\text{HEFT})$ [pb] & $\sigma_{t\times b}$ [pb] & $\sigma_{t\times b} \,(Y_{b,\overline{\text{MS}}})$ [pb] &  $\sigma_{t\times b}/\sigma_\text{HEFT}$ [\%]  \\
\hline
\hline
\multicolumn{6}{c}{$\sqrt{s}=7$~TeV}\\
\hline
$\mathcal{O}(\alpha_s^2)$ & $+5.85$ & -- & $-0.708$ & $-0.439$ & \\
LO & $5.85^{+1.56}_{-1.11}$ & -- & $-0.708^{+0.13}_{-0.19}$ & $-0.439^{+0.10}_{-0.16}$ & $-12$ \\
\hline
$\mathcal{O}(\alpha_s^3)$ & $+7.14$ & $-0.0604$ & $-0.226$ & $-0.264$ & \\
NLO & $12.99^{+2.89}_{-2.14}$ & $-0.0604^{+0.021}_{-0.037}$ & $-0.934^{+0.09}_{-0.07}$ & $-0.703^{+0.11}_{-0.12}$ & $-7.2^{+1.0}_{-0.8}$ \\
\hline
$\mathcal{O}(\alpha_s^4)$ & $+3.28$ & $+0.0386(2)$ & $+0.121(3)$ & $-0.026(2)$ & \\
NNLO & $16.27^{+1.45}_{-1.61}$ & $-0.0218(2)^{+0.035}_{-0.009}$ & $-0.813(3)^{+0.10}_{-0.04}$ & $-0.729(2)^{+0.04}_{-0.01}$ & $-5.0^{+1.0}_{-0.8}$ \\
\hline
\hline
\multicolumn{6}{c}{$\sqrt{s}=8$~TeV}\\
\hline
$\mathcal{O}(\alpha_s^2)$ & $+7.39$ & -- & $-0.895$ & $-0.554$ & \\
LO & $7.39^{+1.98}_{-1.40}$ & -- & $-0.895^{+0.17}_{-0.24}$ & $-0.554^{+0.13}_{-0.20}$ & $-12$ \\
\hline
$\mathcal{O}(\alpha_s^3)$ & $+9.14$ & $-0.0873$ & $-0.268$ & $-0.323$ & \\
NLO & $16.53^{+3.63}_{-2.73}$ & $-0.0873^{+0.030}_{-0.052}$ & $-1.163^{+0.10}_{-0.08}$ & $-0.877^{+0.13}_{-0.14}$ & $-7.0^{+1.0}_{-0.8}$ \\
\hline
$\mathcal{O}(\alpha_s^4)$ & $+4.19$ & $+0.0523(2)$ & $+0.167(3)$ & $-0.022(2)$ & \\
NNLO & $20.72^{+1.84}_{-2.06}$ & $-0.0350(2)^{+0.048}_{-0.013}$ & $-0.996(3)^{+0.12}_{-0.05}$ & $-0.899(2)^{+0.04}_{-0.02}$ & $-4.8^{+0.9}_{-0.8}$ \\
\hline
\hline
\multicolumn{6}{c}{$\sqrt{s}=13$~TeV}\\
\hline
$\mathcal{O}(\alpha_s^2)$ & $+16.30$ & -- & $-1.975$ & $-1.223$ & \\
LO & $16.30^{+4.36}_{-3.10}$ & -- & $-1.98^{+0.38}_{-0.53}$ & $-1.22^{+0.29}_{-0.44}$ & $-12$ \\
\hline
$\mathcal{O}(\alpha_s^3)$ & $+21.14$ & $-0.303$ & $-0.446(1)$ & $-0.623(1)$ & \\
NLO & $37.44^{+8.42}_{-6.29}$ & $-0.303^{+0.10}_{-0.17}$ & $-2.42^{+0.19}_{-0.12}$ & $-1.85^{+0.26}_{-0.26}$ & $-6.5^{+0.9}_{-0.8}$ \\
\hline
$\mathcal{O}(\alpha_s^4)$ & $+9.72$ & $+0.147(1)$ & $+0.434(8)$ & $+0.019(5)$ & \\
NNLO & $47.16^{+4.21}_{-4.77}$ & $-0.156(1)^{+0.13}_{-0.03}$ & $-1.99(1)^{+0.30}_{-0.15}$ & $-1.83(1)^{+0.08}_{-0.03}$ & $-4.2^{+0.9}_{-0.8}$ \\
\hline
\hline
\multicolumn{6}{c}{$\sqrt{s}=13.6$~TeV}\\
\hline
$\mathcal{O}(\alpha_s^2)$ & $+17.47$ & -- & $-2.117$ & $-1.311$ & \\
LO & $17.47^{+4.67}_{-3.32}$ & -- & $-2.12^{+0.40}_{-0.57}$ & $-1.31^{+0.31}_{-0.47}$ & $-12$ \\
\hline
$\mathcal{O}(\alpha_s^3)$ & $+22.76$ & $-0.338$ & $-0.464(1)$ & $-0.659(1)$ & \\
NLO & $40.23^{+9.07}_{-6.77}$ & $-0.338^{+0.11}_{-0.18}$ & $-2.58^{+0.20}_{-0.12}$ & $-1.97^{+0.28}_{-0.28}$ & $-6.4^{+0.9}_{-0.8}$ \\
\hline
$\mathcal{O}(\alpha_s^4)$ & $+10.47$ & $+0.162(1)$ & $+0.464(9)$ & $+0.022(6)$ & \\
NNLO & $50.70^{+4.53}_{-5.14}$ & $-0.176(1)^{+0.14}_{-0.03}$ & $-2.12(1)^{+0.33}_{-0.16}$ & $-1.95(1)^{+0.09}_{-0.03}$ & $-4.2^{+0.9}_{-0.8}$ \\
\hline
\hline
\multicolumn{6}{c}{$\sqrt{s}=14$~TeV}\\
\hline
$\mathcal{O}(\alpha_s^2)$ & $+18.26$ & -- & $-2.213$ & $-1.370$ & \\
LO & $18.26^{+4.88}_{-3.47}$ & -- & $-2.21^{+0.42}_{-0.59}$ & $-1.37^{+0.32}_{-0.49}$ & $-12$ \\
\hline
$\mathcal{O}(\alpha_s^3)$ & $+23.86$ & $-0.362$ & $-0.475(1)$ & $-0.682(1)$ & \\
NLO & $42.12^{+9.51}_{-7.10}$ & $-0.362^{+0.12}_{-0.20}$ & $-2.69^{+0.21}_{-0.13}$ & $-2.05^{+0.29}_{-0.29}$ & $-6.4^{+0.9}_{-0.8}$ \\
\hline
$\mathcal{O}(\alpha_s^4)$ & $+10.98$ & $+0.171(1)$ & $+0.488(9)$ & $+0.027(6)$ & \\
NNLO & $53.10^{+4.75}_{-5.39}$ & $-0.191(1)^{+0.15}_{-0.04}$ & $-2.20(1)^{+0.34}_{-0.17}$ & $-2.03(1)^{+0.09}_{-0.03}$ & $-4.1^{+0.9}_{-0.8}$ \\
\end{tabular}
\end{ruledtabular}
\end{table*}
Table~\ref{tab:results} presents the key findings of this publication. For completeness, we reproduce the HEFT results and the effects of a finite top-quark mass, whose value is taken as $m_t=173.055$~GeV, as in Ref.~\cite{Czakon:2021yub}. Additionally, we provide scale uncertainties, which, in the case of the NNLO top-quark mass effects, constitute an additional novel result of this work. The main result is the top-bottom interference effect $\sigma_{t\times b}$ at the same top-quark mass and a bottom-quark mass of $m_b=4.779$~GeV, also including scale uncertainties. A calculation at a lower value, $m_t=170.98$~GeV, yields a $\sigma_{t\times b}$ compatible with the result given in Table~\ref{tab:results} within the Monte Carlo uncertainty. For $\sigma_t-\sigma_\text{HEFT}$ at 13~TeV, the $\mathcal{O}(\alpha_s^4)$ contribution changes slightly from $0.147(1)$~pb to $0.151(1)$~pb, which is compensated by a shift at $\mathcal{O}(\alpha_s^3)$, leading to a very small change of the NNLO result from $-0.156(1)$~pb to $-0.159(1)$~pb. Thus, we conclude that the effect of small variations of the top-quark mass value is too small to be sensibly quantified. While the size of the NNLO correction to $\sigma_{t\times b}$ is as estimated in Refs.~\cite{LHCHiggsCrossSectionWorkingGroup:2016ypw, Anastasiou:2016cez}, our result exhibits some surprising properties. First, one notices that the top-bottom interference has an even larger effect than the exact top-mass dependence compared to HEFT. This is, however, not too surprising because the HEFT values already account for some of the top-mass effects with the rescaling procedure. Second, the $\mathcal{O}(\alpha_s^4)$ corrections have opposite sign and similar magnitude compared to the $\mathcal{O}(\alpha_s^3)$ contribution, leading to an almost complete cancellation. Furthermore, they are more than twice as large as expected from the NLO scale uncertainties. Together with the fact that we observe an increased scale dependence at NNLO compared to NLO, we can conclude that for top-bottom interference effects in Higgs production, standard scale variation at NLO underestimates the magnitude of higher-order contributions. 

Finally, our result for $\sigma_{t\times b}$ at 13~TeV can be compared to Ref.~\cite{Anastasiou:2020vkr}, where the authors calculate it, albeit at a very large top-quark mass value using the $\overline{\text{MS}}$ scheme for the bottom-quark mass and Yukawa coupling, using a next-to-leading logarithmic approximation and obtain $-1.42$~pb at LO, $-2.05$~pb at NLO, and $-2.18$~pb at NNLO. Since their NNLO value lies almost within our scale uncertainty, we can confirm the quality of their approximation. Additionally, we observe a compelling reduction of the renormalization-scheme dependence.

To further investigate convergence and scheme dependence, we transformed our results to a mixed renormalization scheme, where the bottom mass stays on shell, while the Yukawa coupling $Y_b=\sqrt{2}\,m_b/v$ is renormalized in the $\overline{\text{MS}}$ scheme. Contrary to a complete calculation with both the mass and the Yukawa coupling in the $\overline{\text{MS}}$ scheme, the mixed scheme is straightforward to implement since $\sigma_{t\times b}\propto Y_b$. The Yukawa-coupling running was obtained with \texttt{CRunDec}~\cite{Schmidt:2012az} and the required scheme-translation constants can be found in Ref.~\cite{Gray:1990yh}. The results are shown in Table~\ref{tab:results}. At NNLO, they are compatible with the on-shell results within scale uncertainties, albeit with a much weaker scale dependence. The convergence across different orders of perturbation theory is substantially improved, indicating a potentially better convergence of a consistent $\overline{\text{MS}}$-scheme result. At present, our on-shell results with their larger errors seem appropriate as a conservative estimate for phenomenological applications.

In order to make sure that our calculation is correct and the surprising features listed above are not due to mistakes in the calculation, we have successfully reproduced the fixed-order differential distributions at the central scale presented in Refs.~\cite{Caola:2018zye, Jones:2018hbb}. This serves as a check of the real-virtual and real-real contributions. For checks of the three-loop Higgs-gluon form factor by comparison to previous work \cite{Davies:2019nhm, Harlander:2019ioe}, we refer to Ref.~\cite{Czakon:2020vql}.

\section{Conclusions and Outlook}
We have computed the top-bottom interference contribution to the Higgs production cross section in gluon-gluon fusion at NNLO. The result addresses one of the most important remaining theory uncertainties for this process. We find that the effect is sizable and, therefore, crucial for precision predictions at the percent level. The result is compatible with conservative estimates previously available in the literature. Interestingly, the scale uncertainties grow slightly from NLO to NNLO. It has been noticed \cite{Anastasiou:2016cez, Caola:2018zye} that the interference can be very sensitive to the choice of the renormalization scheme. When changing only the renormalization of the bottom Yukawa coupling to $\overline{\text{MS}}$, compatible results with weaker scale dependence are obtained. Since it is questionable if such a mixed renormalization scheme is consistent in the context of the complete Standard Model, and since the standard in the Higgs Working Group is to renormalize the bottom mass in the $\overline{\text{MS}}$ scheme~\cite{LHCHiggsCrossSectionWorkingGroup:2016ypw}, a calculation in the $\overline{\text{MS}}$ scheme for both $m_b$ and $Y_b$ would be highly desirable. Furthermore, the effect of a finite bottom mass in loops which do not couple to the Higgs, i.e.\ in the 4-flavor scheme, has seen recent attention in an approximation at leading power in $1/m_H$~\cite{Pietrulewicz:2023dxt}. As a first rough assessment of such effects, we observe that contributions including $b$-quark PDFs contribute only 0.037~pb to our $\sigma_{t\times b}$ result at NNLO and 13~TeV. We expect the effect of a switch to the 4-flavor scheme to be of a similar order of magnitude. Still, a complete calculation in this scheme remains of major interest. We defer a dedicated analysis comparing the calculation presented in this letter to an exact calculation in the $\overline{\text{MS}}$ scheme and/or the 4-flavor scheme to future work. Due to the open questions regarding the scheme choice, we refrain from providing values for top-charm interference, even though the methods used in this work can be readily applied to obtain them. It seems sensible to first address the impact of scheme choice before considering this very small effect, where the usage of the 5-flavor scheme as explained above might no longer be appropriate.

We are grateful to Jonas Lindert for providing us with raw data from Ref.~\cite{Caola:2018zye}, which enabled a check of our calculation. This work was supported by the Deutsche Forschungsgemeinschaft (DFG) under grant 396021762 -
TRR 257: Particle Physics Phenomenology after the Higgs Discovery, and grant 400140256 - GRK
2497: The Physics of the Heaviest Particles at the LHC. The authors gratefully acknowledge the computing time provided to them at the NHR Center NHR4CES at RWTH Aachen University (project number p0020025). This is funded by the Federal Ministry of Education and Research, and the state governments participating on the basis of the resolutions of the GWK for national high performance computing at universities (\url{www.nhr-verein.de/unsere-partner}).

\bibliography{main}

\end{document}